\documentclass[final,1p,times]{elsarticle}
\biboptions{longnamesfirst,compress}

\usepackage{amsmath}
\usepackage{amssymb}
\usepackage{siunitx}
\usepackage{graphicx}
\usepackage{epstopdf}
\usepackage{chngcntr}
\usepackage{subfigure}
\usepackage{enumerate}
\usepackage[nomarkers,figuresonly]{endfloat}
\usepackage{booktabs}
\usepackage{multirow}
\usepackage{xcolor}
\renewcommand{\footnotesize}{\scriptsize}
\counterwithout{figure}{section}

\journal{Journal of Nuclear Material, SI: MPIM 2015}

\begin{document}

\begin{frontmatter}



\title{Study of the Mechanical Behavior of the Hydride Blister/Rim Structure in Zircaloy-4 using \emph{in-situ} Synchrotron X-ray Diffraction}


\author[UIUC]{Jun-li Lin}
\author[UIUC]{Xiaochun Han}
\author[UIUC]{Brent J. Heuser\corref{cor1}}
\author[APS]{Jonathan D. Almer}
\cortext[cor1]{Corresponding author: Tel: +1-217-333-9610, email: bheuser@illinois.edu}
\address[UIUC]{Department of Nuclear, Plasma, and Radiological Engineering, University of Illinois at Urbana-Champaign, Urbana, IL, USA}
\address[APS]{Advanced Photon Source, Argonne National Laboratory, Argonne, IL, USA}
\begin{abstract}
High-energy synchrotron X-ray diffraction was utilized to study the mechanical response of the f.c.c $\delta$ hydride phase, the intermetallic precipitation with hexagonal \emph{C14} lave phase and the $\alpha$-Zr phase in the Zircaloy-4 materials with a hydride rim/blister structure near one surface of the material during \emph{in-situ} uniaxial tension experiment at 200 \si{\celsius}. The f.c.c $\delta$ was the only hydride phase observed in the rim/blister structure.  The conventional Rietveld refinement was applied to measure the macro-strain equivalent response of the three phases. Two regions were delineated in the applied load versus lattice strain measurement: a linear elastic strain region and region that exhibited load partitioning. Load partitioning was quantified by von Mises analysis. The three phases were observed to have similar elastic modulus at 200 \si{\celsius}.
\end{abstract}


\end{frontmatter}


\section{Introduction}
Zircaloy-4 is a common fuel cladding used in pressurized light water nuclear reactors due to excellent mechanical properties, corrosion resistance, and low thermal neutron absorption cross-section. However, in-service absorption of hydrogen by Zircaloy-4 (Zry-4) is potentially problematic.  One phenomenon of potential importance is delayed hydride cracking (DHC) which can occur during storage of used nuclear fuel (UNF).  DHC is a time-dependent subcritical crack growth mechanism that can be induced by tensile hoop stress within the clad wall below the yield stress \cite{puls1990effects,kim2008delayed}.  DHC can eventually lead to clad failure during dry storage, potentially resulting in the release of radioisotopes within the storage casks.

The root-cause of the DHC is hydride re-orientation, where, under sufficient stress, plate-like hydrides dissolve and  preferentially reprecipitate with plate normal axis parallel to the stress axis \cite{kearns1966effect,coleman1982effect}. This potential issue has motivated studies of the microstructure and mechanical response of the hydride phase under externally applied tensile stress to understand the reorientation behavior.  For example, the threshold stress of the hydride reorientation is observed to be affected by several factors, including fabrication history of the material \cite{kearns1966effect}, strength and grain size of the matrix \cite{kearns1966effect,leger1985effect}, and the testing temperature \cite{singh2004stress,bell1975hydride,hong2005stress}.  For the typical cold-worked stress relief (CWSR) Zry-4, the threshold stress of the hydride re-orientation is observed $\sim$80 MPa at 400\si{\degreeCelsius} \cite{daum2006radial,colas2010situ}.

High energy synchrotron X-ray has been widely utilized to study the characteristics of the minor hydride phase in zirconium based material under \emph{in-situ} load-temperature conditions.  Taking advantage of the penetrability of the high-energy incident X-ray beam in transmission geometry, the precipitation of hydride phase in zirconium based materials was identified \cite{daum2009identification,tulk2012study}. The dissolution and precipitation temperature of $\delta$ hydride phase in Zry-2 and Zry-4 were measured \cite{colas2010situ,zanellato2012synchrotron}.  The texture of the $\delta$ hydride phase was also studied \cite{santisteban2010hydride}.  Kerr \emph{et al.} \cite{kerr2008strain} studied the lattice strain evolution of zirconium and $\delta$-hydride phase as the function of applied stress in samples with a uniform distribution hydride.  These authors observed a load partitioning behavior between the matrix and the $\delta$-hydride phase under an uniaxial tension test.  The load was observed transfer to the hydride phase from the soft zirconium matrix phase at an applied stress of 267 MPa at the room temperature; with tensile stress applied parallel to the transverse direction.

Synchrotron X-ray diffraction was also used to study hydride reorientation, with two different diffraction based signatures applied to study the reorientation behavior \cite{colas2010situ,colas2013effect,alvarez2012hydride}.  First, Colas \emph{et al.} \cite{colas2010situ,colas2013effect} studied the evolution of the full-width at half maximum (FWHM) of the $\delta(111)$ hydride peak as samples were cooled to induce hydride precipitation with or without the stress.  An increase of the FWHM was observed for radial hydride particles with the plate normal oriented along the direction of the applied load, while for circumferential hydride the FWHM was observed invariant in the same direction during the cooling.  This increase of FWHM was attributed to strain broadening with minor or no contribution from the particle size broadening.  Second, Alvarez \emph{et al.} \cite{alvarez2012hydride} proposed that the sign of hydride reorientation could be obtained from the interplanar spacing of $\delta(111)$ reflection based on the knowledge that the circumferential and radial hydride platelets have different preferential growth directions.

While most of the literature have results focused on materials with uniformly distributed hydride phase, the hydride rim (a near-surface layer contained high concentration of hydrides) and hydride blister structure (a layer contained very high concentration of hydrides or bulky solid hydride phase) is typically observed on the water-side of post-service cladding.  This hydride morphology is generated by the temperature gradient within cladding wall  \cite{lemaignan1994zirconium,sawatzky1985formation,garde1996effects}.  The hydride rim/blister morphology causes a significant loss of ductility of the Zr-based matrix \cite{daum2002embrittlement}. The brittle fracture characteristics of a hydride blister was also observed by Pierron \emph{et al.} \cite{pierron2003influence} at a temperature range of 25 to 400 \si{\degreeCelsius}.   The texture of the $\delta$ hydride phase in the blister morphology was observed has the same relationship with the $\alpha$-Zr ($\alpha(0001)//\delta(111)$) as the uniform distributed hydride phase does \cite{alvarez2011phase}.  Daum \emph{et al.} \cite{daum2009identification} studied the hydride phase with rim morphology in Zry-4 cladding by synchrotron X-ray diffraction at the room temperature. The $\delta$ hydride was the only hydride phase identified with overall hydrogen contents $\leq$1250 wt.ppm in these studies.  These authors also observed that the formation of the rim in Zry-4 cladding did not cause the measurable diffraction peak shift of the $\alpha$-Zr phase.

The present work aims to examine the lattice strain response of the hydride and the second intermetallic phase in CWSR Zry-4 material with a near-surface rim/blister structure under an applied tensile load at 200 \si{\celsius}, applying the Rietveld method to measure the macro-strain equivalent response of each phases. The internal stress on each phases were then quantified by von Mises effective stress. The results aim for providing significant data for future works to better predict the claddings performance during the fuel cycle. Short-term creep testing (5 to 6 hrs at 200 \si{\celsius}) was also performed to simulate the dry storage conditions.

%
\section{Experimental}
\subsection{As-received material}
\begin{figure}[!ht]
\centering
\subfigure[] {\label{fig:pole_figure}
\includegraphics[scale=0.5]{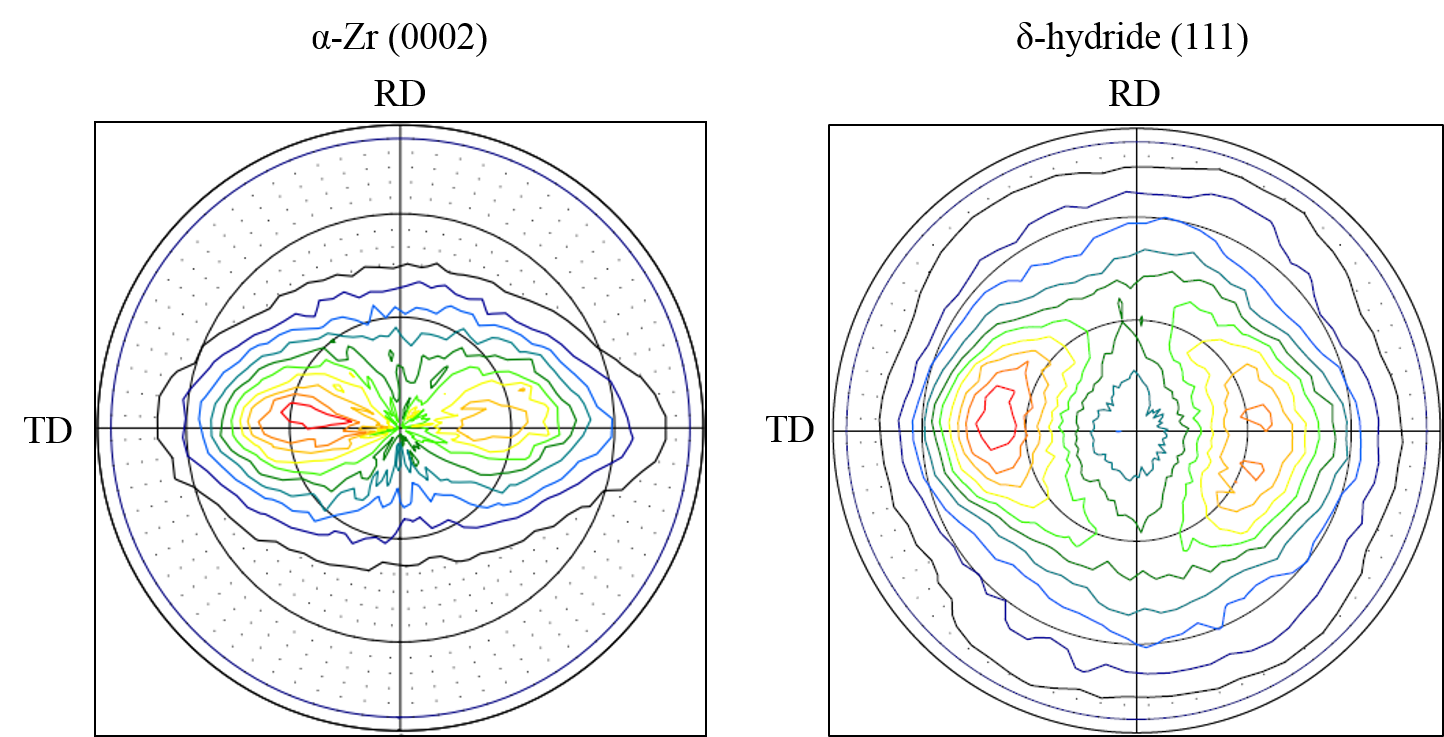}}
\\
\subfigure[] {\label{fig:dogbone}
\includegraphics[scale=0.6]{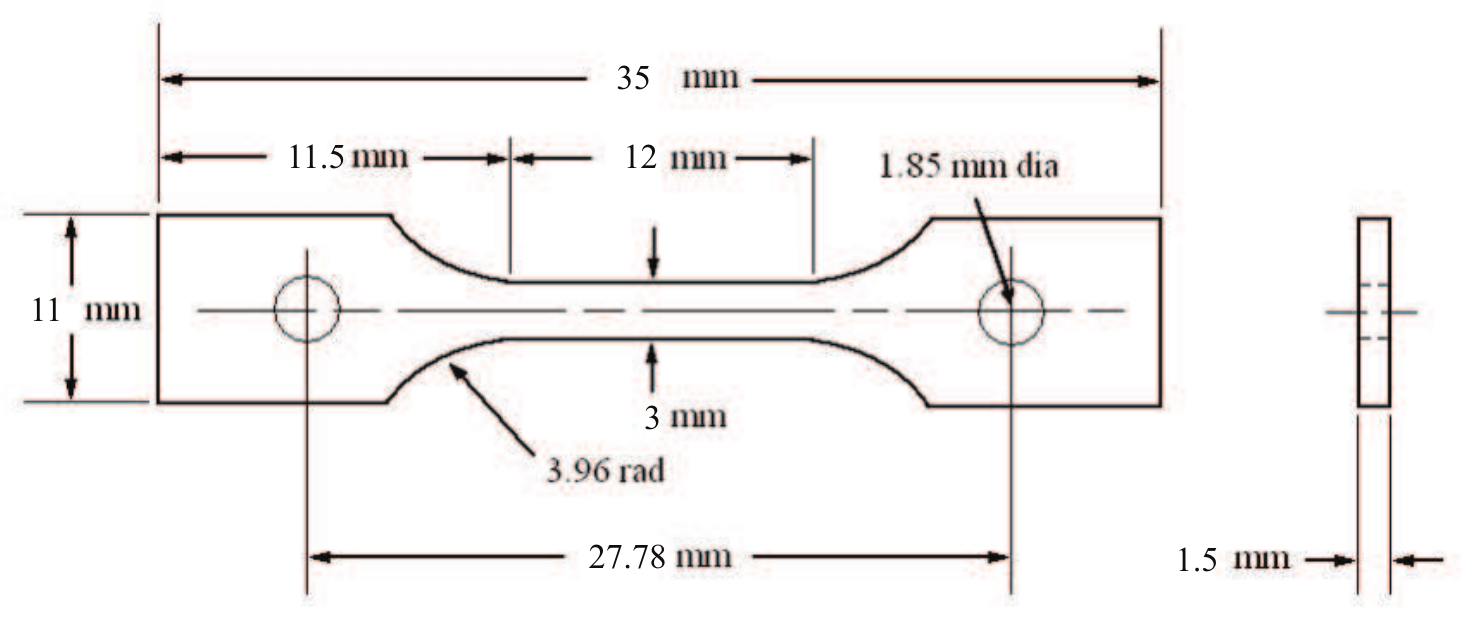}}
\caption{(a) (0002) basal pole figure for the as-received Zry-4 sheet. (b) The dimension of the tensile dog-bone specimens. Load will be applied along the RD during the uniaxial tensile test.}
\end{figure}

A CWSR Zircaloy-4 sheet with thickness of 1.5 mm was obtained from ATI Specialty Alloys and Components.  The chemical composition of this material was very similar to what is specified in ASTM Standards B353-12 \cite{astmb353}.  The texture measurement shown in Fig. \ref{fig:pole_figure} exhibits the typical basal pole figure alignment for CWSR Zry-4 material; the two c-axis poles are tilted $\pm$30 to 40 degree from the normal direction (ND) and resides within the transverse-normal (TD-ND) plane \cite{lemaignan1994zirconium,douglass1971}.  Typical tensile dog-bone specimens were machined from the sheet with a dimension specified in Fig. \ref{fig:dogbone}, with applied load along the rolling direction (RD).  We note that for in-service cladding the tensile hoop stresses act along the TD not the RD.
%
\subsection{Hydrogen charging}
Tensile specimens were machined from the Zry-4 plate by electrical discharge machining (EDM) to avoid generating residual stresses. The specimens were then hydrided at 400 \si{\celsius} using a Sievert-type apparatus gas charging system \cite{chen2000solubility} to produce hydride with rim/blister distribution in the specimens. Specimens were mechanically polished up to 0.05 \si{\micro\meter} using silicon dioxide solution.  A thin film of Ni ($\sim0.2 \si{\micro\meter}$) was sputter coated onto one surface of the specimens to enhance the near-surface hydrogen chemical potential during the charging, thereby achieving the desired rim/blister hydride morphology \cite{pierron2003influence}.  The post-charging procedure was to slowly cool the specimens to  room temperature at rate of 1 \si{\degreeCelsius\per\minute} to ensure the precipitation of the $\delta$-hydride phase \cite{bradbrook1972precipitation}.  The charging was performed under stress-free environment to avoid forming radial hydride. The overall hydrogen content was determined by measuring the reduction of the hydrogen gas pressure using the ideal gas law \cite{chen2000solubility}.

Optical metallography was performed to examine the hydride morphology. The TD-ND plane of samples were polished using the 1200 grits paper and then etched by wiping the plane with an acid solution composed of DI water, nitric acid, sulfuric acid and hydrofluoric acid with ratio of 10:10:10:1 for 10 seconds.  The result of an optical image is shown in Fig. 2. The thickness of the rim/blister structure was estimated as 40 and 20 \si{\micro\meter} for sample with $\sim1000$ and $\sim500$ wt.ppm hydrogen, respectively. For sample with $\sim100$ wt.ppm hydrogen, the rim/blister was less than 5 \si{\micro\meter} and distributed nonuniformly.  Electron backscatter diffraction (EBSD) was also performed to characterize the particle distribution for sample charged with deuterium using the same charging procedure.  A Hitachi IM4000 ion milling system was used to create a damageless cross section for EBSD analysis.  This system utilizes a broad, low-energy Ar+ ion beam to remove materials on the sample surface and can run in both flat milling and cross section milling modes.  An example of EBSD image is shown in Fig. 3 and shows the formation of rim/blister deuteride near the surface of the sample.  We note here because of the destructive nature for both optical and EBSD analyses, the samples shown in Fig. 2 and 3 are not used in this study. However, specimens for this study were prepared using the same procedures as described above and were expected to yield a very similar result.

Both the optical microscopy and the EBSD result demonstrated the presence of the hydride rim/blister near the surface. A rough estimation using image processing tool box in Matlab to measure the fraction of hydride in the rim/blister structure, which yields $\sim25\%$, $\sim10\%$ and less than $1\%$ for samples with $\sim1000$, $\sim500$ and $\sim100$ wt.ppm hydrogen concentration, respectively. These values provide a rough sense of the contribution of hydride rim/blister structure to the total diffraction intensity from the $\delta$-hydride phase.

\begin{figure}[!ht]
\label{fig:om}
\centering
\includegraphics[scale=0.5]{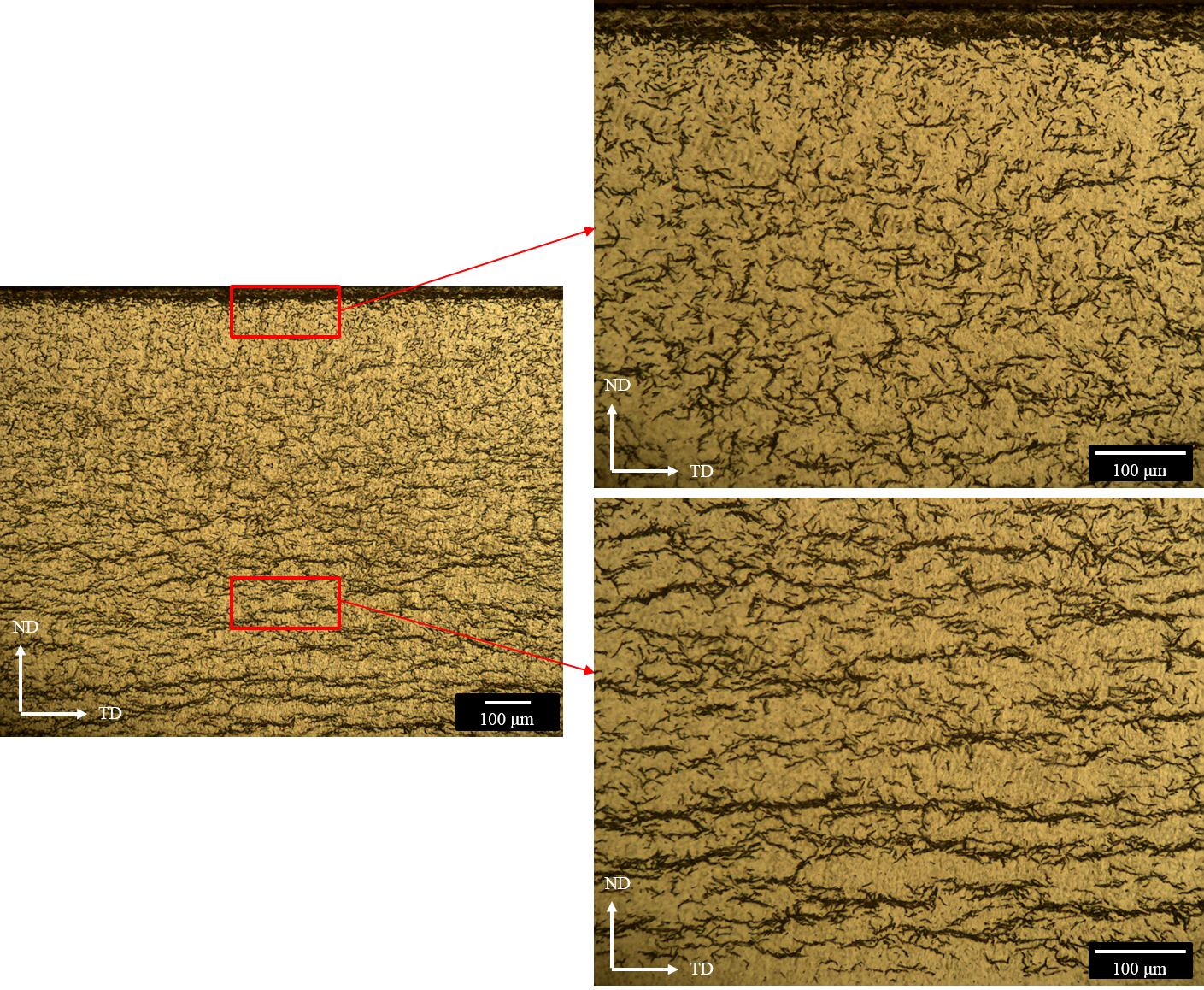}
\caption{A near-surface high-density hydride rim/blister layer was observed. The hydride particles distributed uniformly underneath the rim/blister structure with long axis along the TD. The sample has $\sim1000$ wt. ppm hydrogen.}
\end{figure}

\begin{figure}[!ht]
\label{fig:rim_ebsd}
\centering
\includegraphics[scale=0.5]{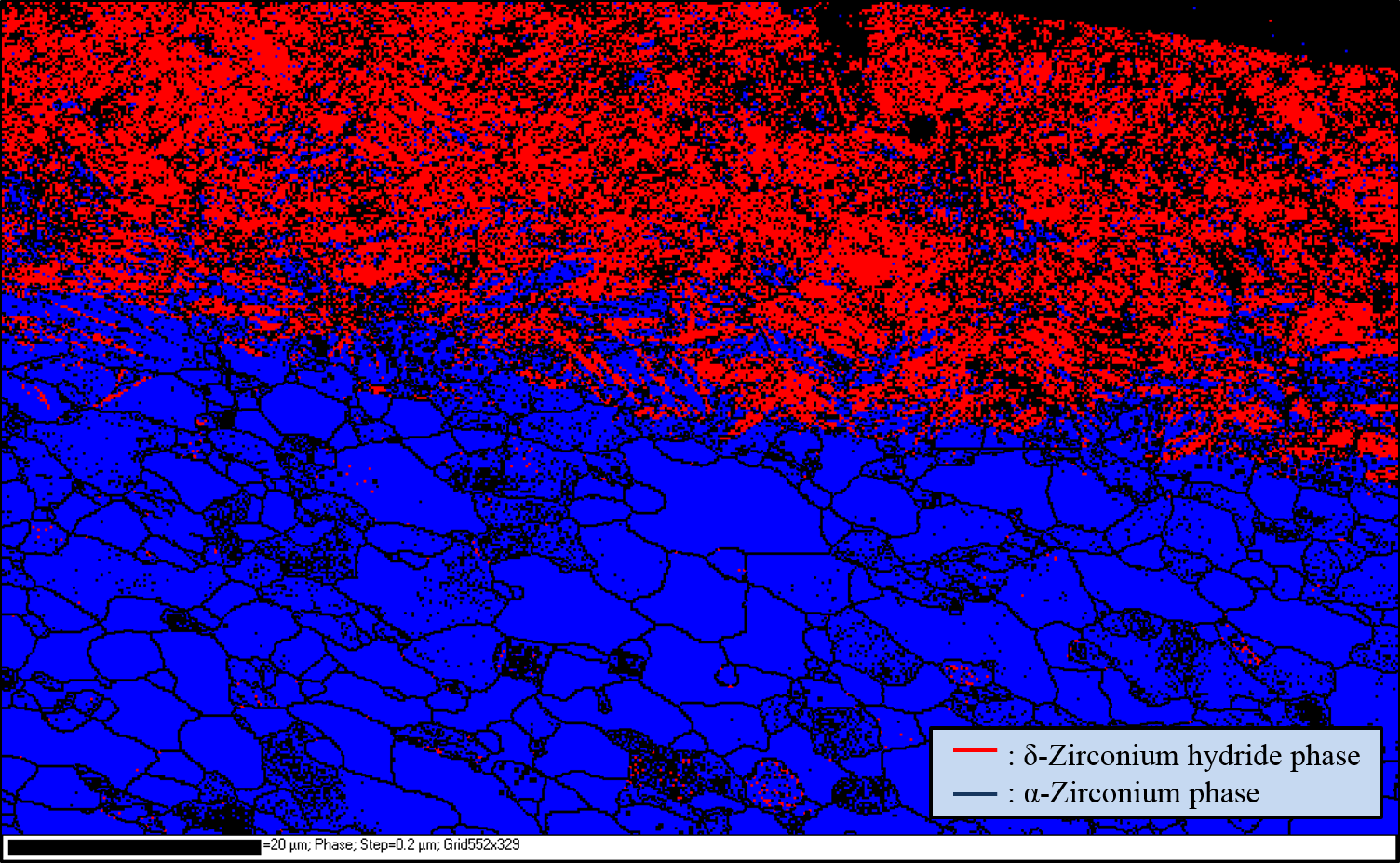}
\caption{An EBSD image for a 1141 wt.ppm deuterium sample. The $\delta$ deuteride phase was identified and marked in red while the blue area indicates the $\alpha$-Zr phase.}
\end{figure}
\subsection{$\it\text{In-situ}$ synchrotron x-ray diffraction}
\begin{figure}[!ht]
\centering
\includegraphics[scale=0.6]{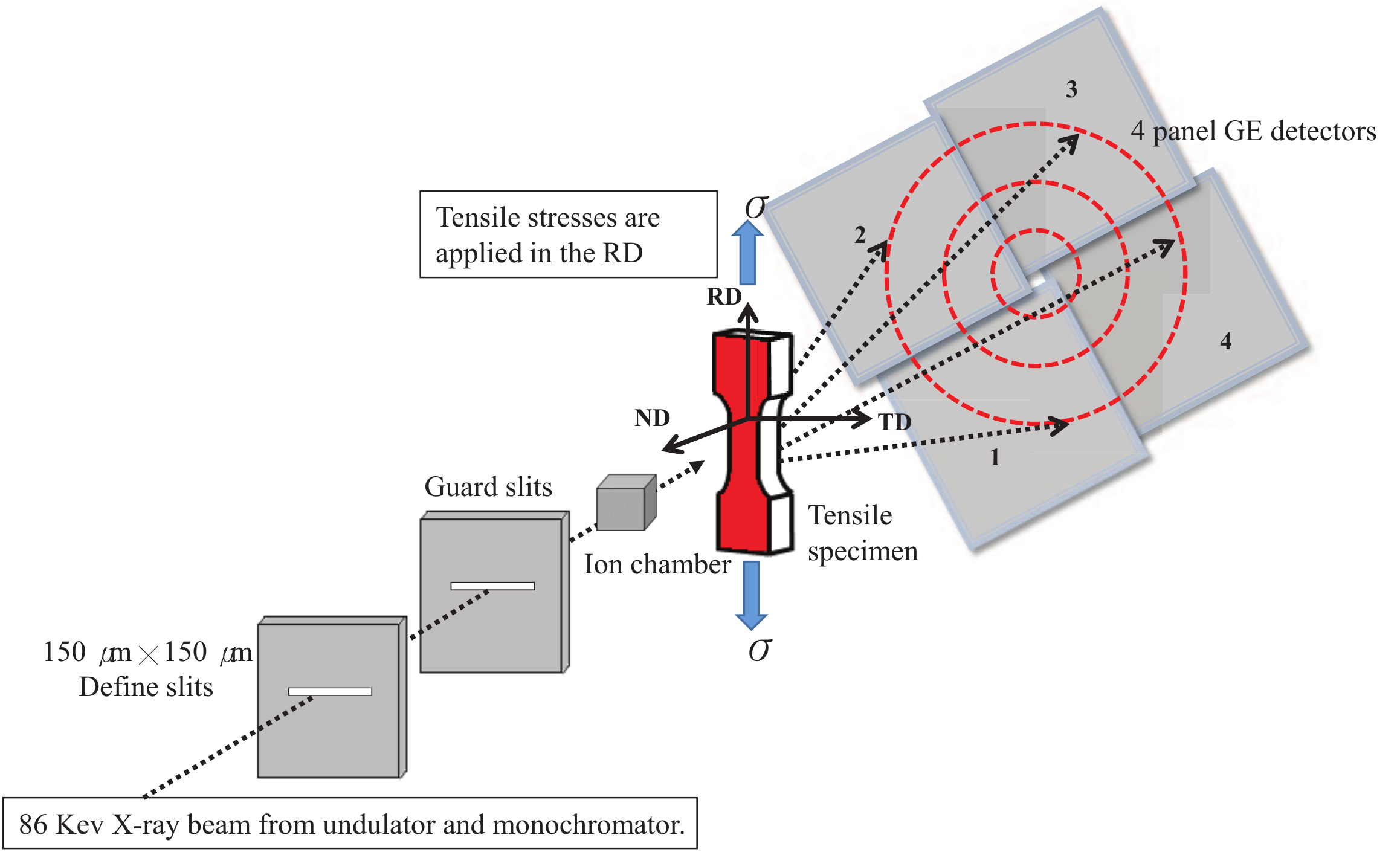}
\caption{Schematic of the 1-ID beamline and the experiment setup at APS. The red area indicates the near-surface area where the hydride rim/blister structure was formed.}
\label{fig:setup1}
\end{figure}

Synchrotron X-ray diffraction analysis was performed at the 1-ID Beamline of the Advanced Photon Source at Argonne National Laboratory \cite{haeffner2005use}.  An incident X-ray beam energy of 86 \si{\kilo\electronvolt} was used with size of 150$\times$150 \si{\micro\meter\squared}.  A set of high-resolution quad GE a-Si detectors (four panels Hydra system) with 2048$\times$2048 pixels per detector was used.  The detector arrangement used for these experiments is shown in Fig. (\ref{fig:setup1}). The sample to detector distance is about 2060 mm. The MTS$^{\textregistered}$ load frame and a clam shell furnace was used for \emph{in-situ} uniaxial tension and heating.  The load was applied along the RD of the samples.
The data acquisition program was set up to automatically scan 12 individual data points in the gauge region of the sample.  These 12 individual data images were summed to increase the intensity of the minor hydride phase and to avoid detector saturation due to much stronger Zr matrix diffraction intensity.  This procedure also effectively increased the measured volume and minimized the sampling effect as population of hydride and matrix phase was anticipated to shift in space under applied load at temperature.

Temperature control was performed with a thermal couple mounted at the bottom of the gauge region out of the incident beam. Presumably the temperature gradient between the thermal couple and the beam measured area at the testing temperature (200 \si{\celsius}) is small.
Samples were first measured at the room temperature with small applied load (10 N) and then heated to 200 \si{\celsius}. The applied load was then increased incrementally to 700 N, with associated tensile stress of $\sim$ 200 MPa.  Table \ref{tab:test_setup} lists the experimental conditions of these measurements.  Samples were then held at 200\si{\degreeCelsius} at the maximum stress for 4 to 8 hours to simulate the short-term creep test.

\begin{table}[!ht]
\begin{center}
\caption{Summary of the thermal-mechanical procedures used for the samples in this study.}
\begin{tabular}{cccc}\hline\hline
\textbf{[H] content}&\textbf{Max Temp.}&\textbf{$\sigma_{max}$}&\textbf{Holding period}\\
\textbf{(wt.ppm)}&\textbf{(\si{\degreeCelsius})}&\textbf{(\si{\mega\pascal})}&\textbf{(hr)}\\\hline
977&200&218&4.5 \\
492&200&200&3.5 \\
98&200&190&4 \\
0&Room temperature&None&None\\ \hline\hline
\end{tabular}
\label{tab:test_setup}
\end{center}
\end{table}
\subsection{Data analysis}
\begin{figure}[!ht]
\centering
\subfigure[] {\label{fig:ring}
\includegraphics[scale=0.5]{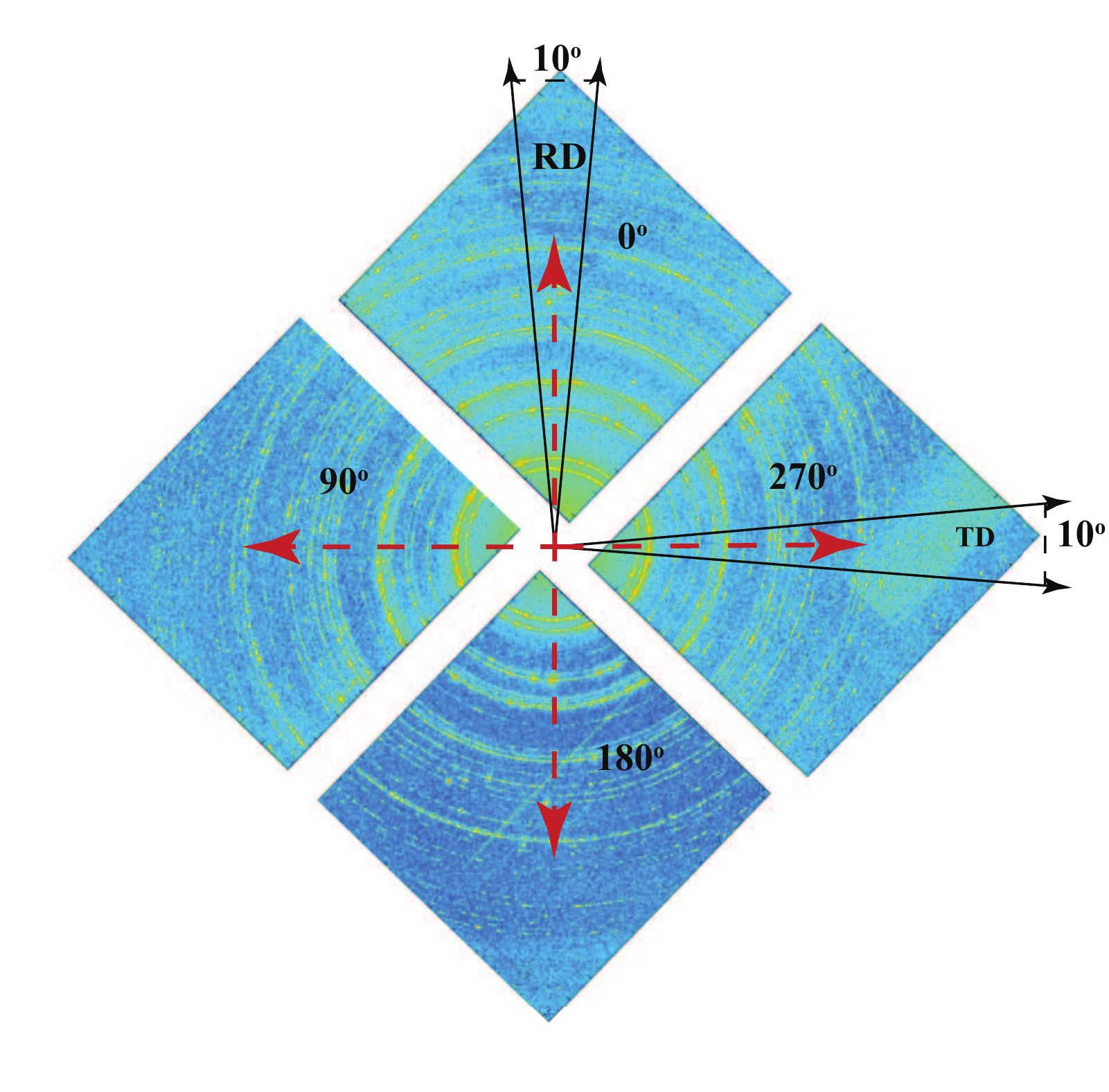}}
\\
\subfigure[] {\label{fig:trans_fig}
\includegraphics[scale=0.5]{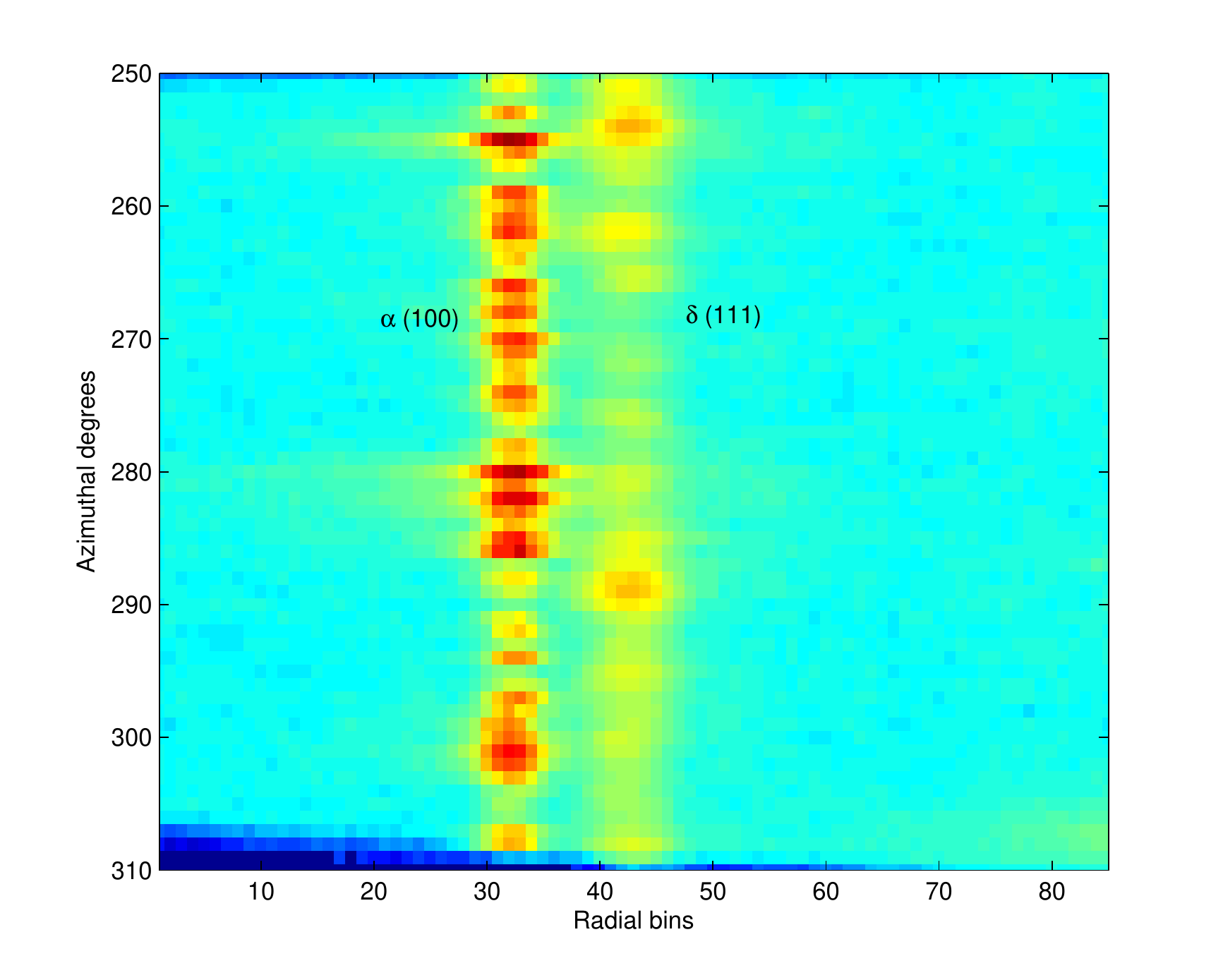}}
\caption{An example of diffraction image for the 977 wt.ppm sample. (a) A typical Debye Sherrer ring composed by four detectors and (b) a re-bin image of the  reflection intensity along the azimuthal angle versus the radial distance, showing the $\delta(111)$ hydride and $\alpha(100)$ zirconium relfection. }
\end{figure}

The 1-ID instrument was calibrated by Fit2D program by measuring the diffraction pattern from cerium oxide \cite{hammersley1994calibration} under the same experiment geometry, yielding the detector tilts and positions relative to the incident X-ray beam. The information was then input in Matlab$^\text{\textregistered}$ routine for further analysis such as diffraction image processing and background subtraction \cite{almer2003strain}.
An example of the four panel detector raw data output showing multiple Debye-Scherer rings is shown in Fig. \ref{fig:ring}.  These data can be re-binned to show diffraction peak or reflection intensity along the azimuthal angle about the Debye-Scherer ring(s) versus the radial distance, the latter related to the inter-planar spacing via the Bragg diffraction law. Such a re-binning is shown in Fig. \ref{fig:trans_fig} for panel detector 2.  This image is for the 977 wt. ppm sample and shows the $\delta$ hydride phase and the $\alpha$ zirconium phase diffraction intensity.  This type of data can be further analyzed to determine phase orientation relationships and texture using, for example, MTEX \cite{hielscher2008novel} or E-WIMV algorithm \cite{santisteban2010hydride}).  The pole figure analysis on the $\delta$ hydride (111) reflection (Fig. 1(a)) suggests that the texture of the hydride phase and the relationship with the $\alpha$-zirconium is not altered from the known properties of the Zr-hydride system, which most of $\delta$(111) planes are parallel to the $\alpha$-Zr basal plane, by the rim/blister morphology \cite{alvarez2011phase} as previously mentioned.

Diffraction intensities were extracted with respect to four principal directions on the four detectors: 0, 90, 180, and 270 degrees, integrating $\pm$5 degrees around these directions. The applied load (RD) and the orthogonal (TD) directions are defined along 0/180 and 90/270 degrees, respectively as shown in Fig. \ref{fig:ring}. The TD diffraction data represents the Poisson's response to the applied tensile stress. The processed data was then exported into GSAS friendly format in order to measure lattice parameters using the conventional Reitveld refinement method \cite{larson1994gsas,toby2001expgui}. CMPR was also used to fit the initial peak profile for the refinement \cite{toby2005cmpr}.

\section{Result and discussion}
\subsection{Phase identification}
An example of Reitveld refinement for the 977 wt.ppm sample is shown in Fig. 6. Phases were identified with the aid of standard powder diffraction data (PDF) \cite{pdfcard}. $\alpha$-Zr is observed as the dominant phase. The $\delta$-hydride phase is the only hydride phase observed and the strongest diffracted intensity corresponds to (111) reflection, which is in good agreement within the known hydride phases formed in CWSR Zry-4 for hydrogen concentrations $\leq$1250 wt.ppm with the rim morphology \cite{daum2009identification}.
Diffraction intensity from the so-called $\chi$ phases or hexagonal C14 laves phase with composition of Zr(Fe$_{1.5}$Cr$_{0.5}$)$_2$ \cite{erwin2001observation} was also observed.
\begin{figure}[!ht]
\begin{center}
\subfigure[] {\label{fig:lineout_1}
\includegraphics[scale=0.4]{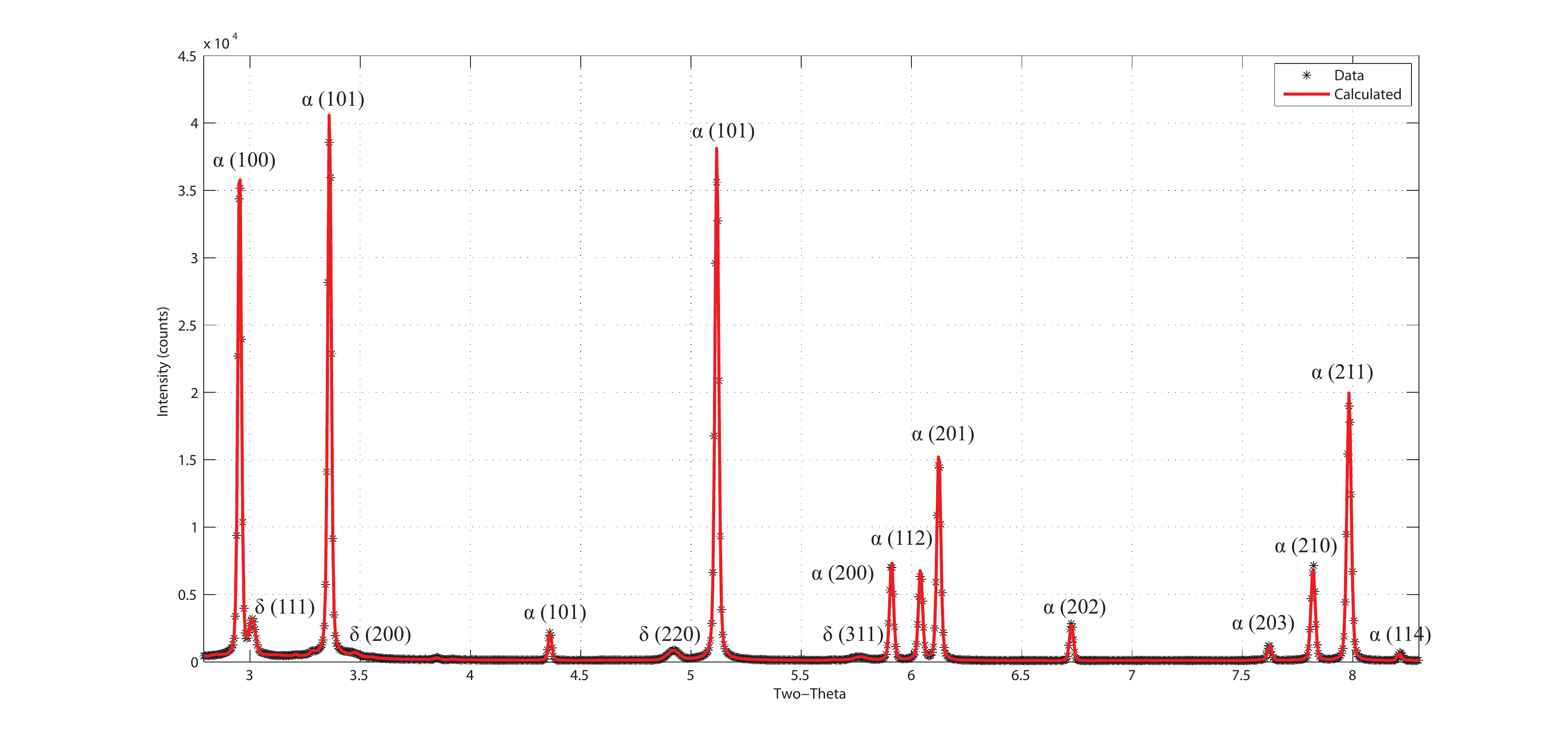}}
\\
\subfigure[] {\label{fig:delta111_line}
\includegraphics[scale=0.27]{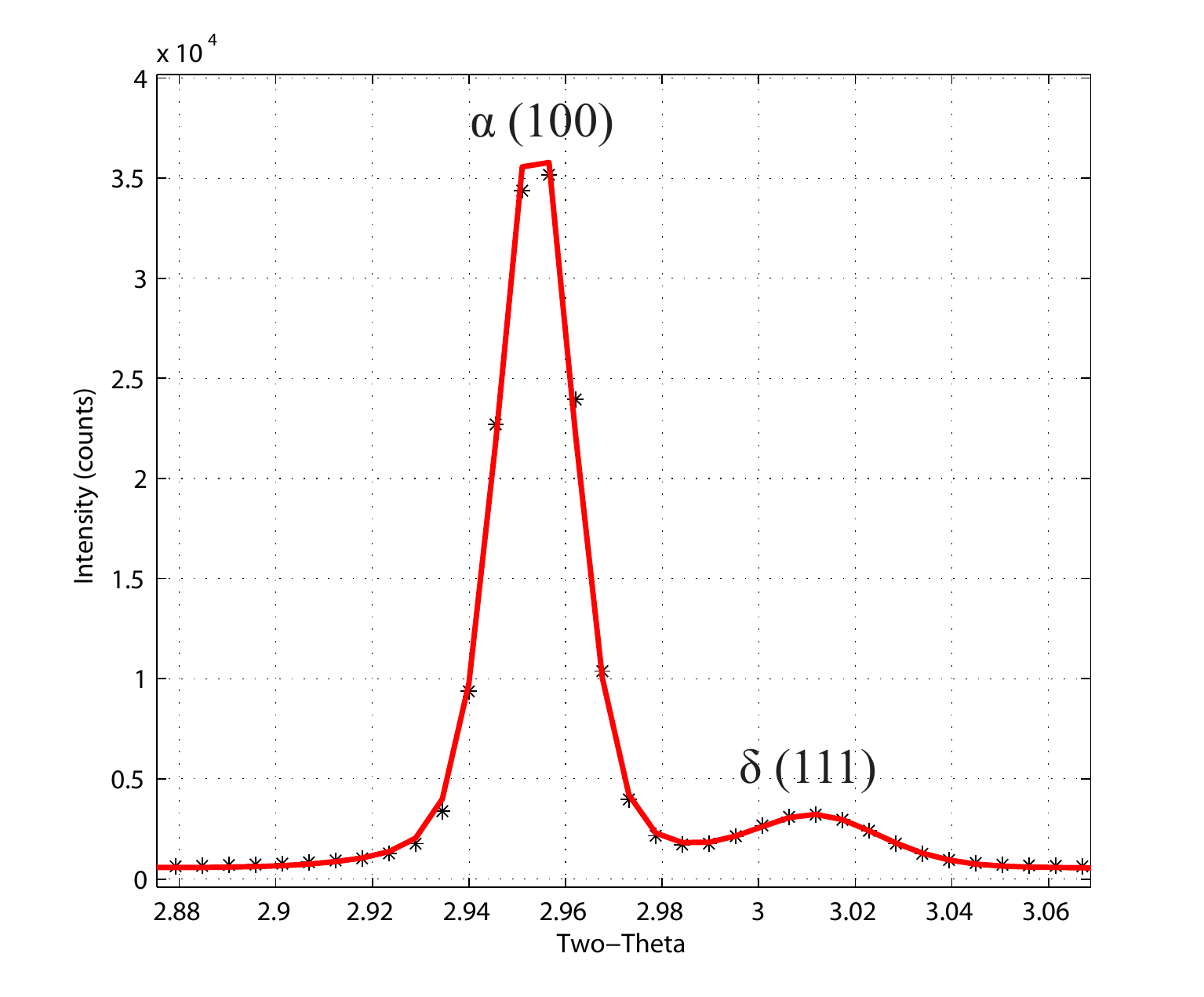}}
\subfigure[] {\label{fig:fit_chi}
\includegraphics[scale=0.27]{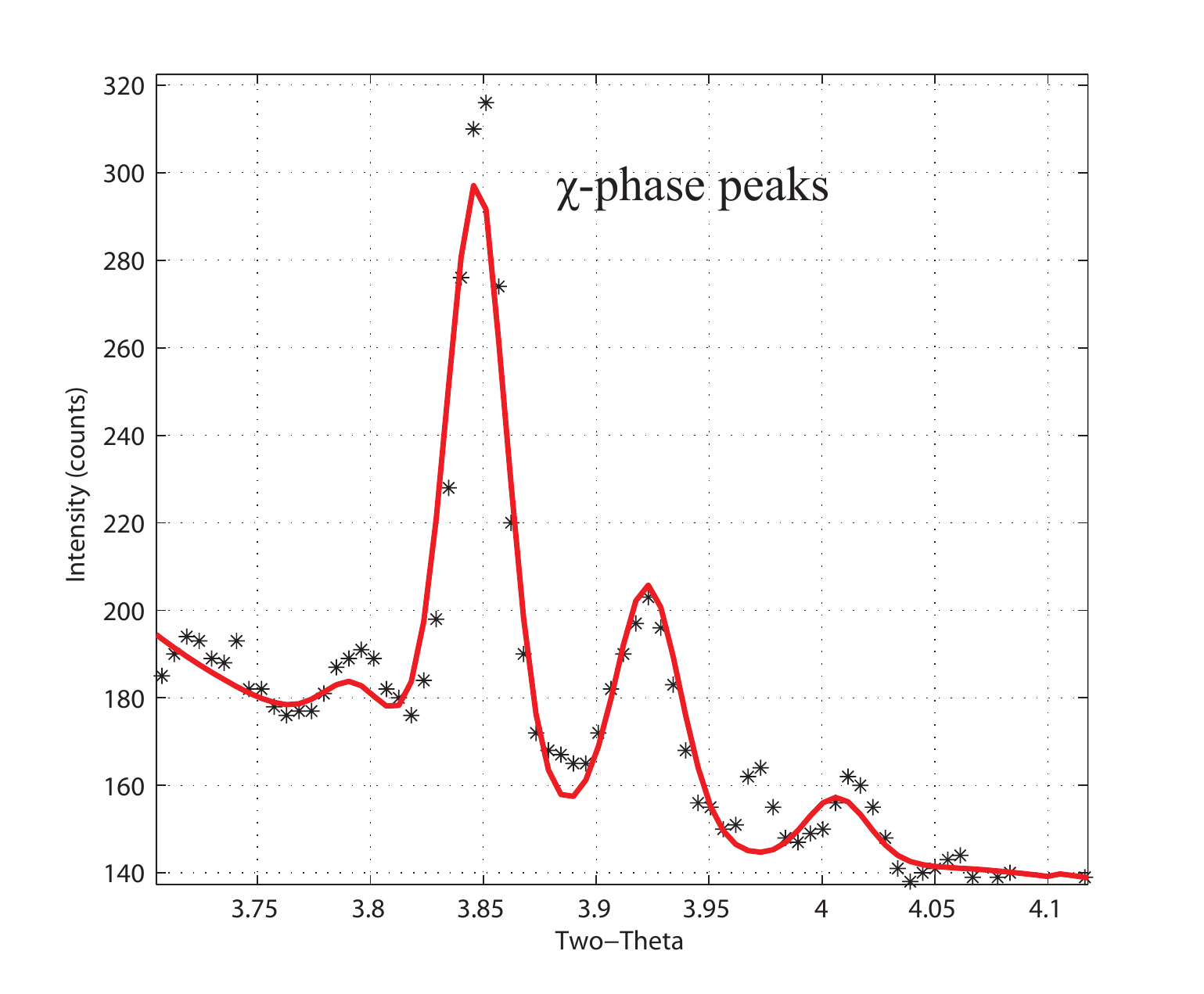}}
\subfigure[] {\label{fig:fit_tail}
\includegraphics[scale=0.27]{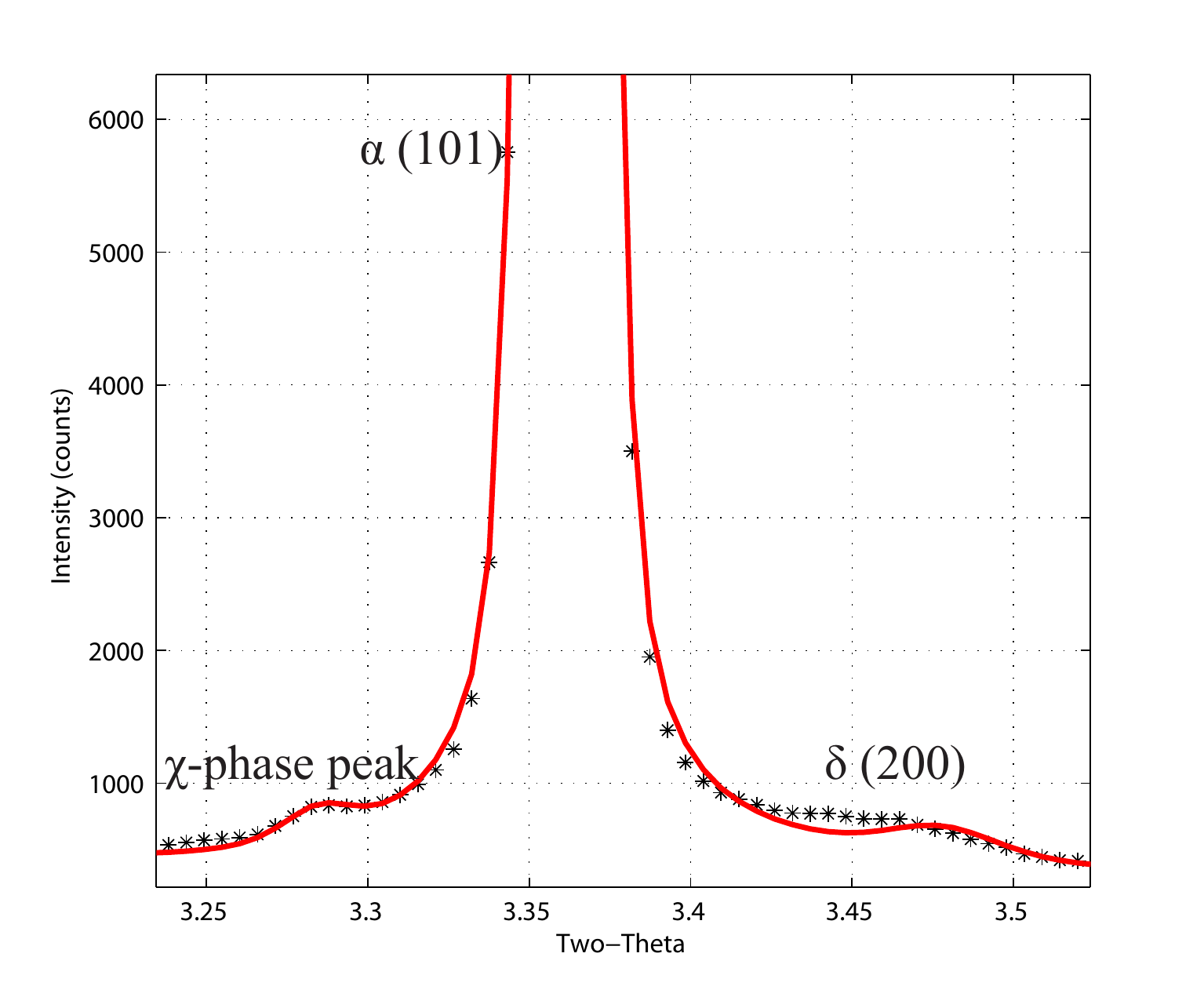}}
\caption{An example of Reitveld refinement fit in the RD segment for the 977 wt.ppm sample, showing the resolvable hydride and $\chi$-phase peaks as well as the refinement result. Data here was taken at room temperature without load.}
\end{center}
\end{figure}
%

The lattice constants of the three phases obtained from the Reitveld method were summarized in Table \ref{tab:lattice_const}, which were measured at room temperature without load.
\begin{table}[b]
\begin{center}
\caption{Summary of lattice constants of the three phases.}
\label{tab:lattice_const}
\begin{tabular}{ccccccc}
  \toprule
  H Content & Direction & \multicolumn{2}{c}{$\alpha$-Zr} & \multicolumn{2}{c}{$\chi$-phase} & $\delta$-hydride \\
  \cmidrule{3-4}
  \cmidrule{5-6}
  (Wt.ppm)       &       & a$_0$ ($\AA$) & c$_0$ ($\AA$) & a$_0$ ($\AA$) &   c$_0$ ($\AA$)    &       a$_0$ ($\AA$)   \\
  \midrule
  \multirow{2}{*}{None}  & RD & 3.238 & 5.167   & 5.022    & 8.359  & - \\
                         & TD & 3.237 & 5.167   & 5.039    & 8.287 & -  \\
  \multirow{2}{*}{98}  & RD & 3.239 & 5.172   & 5.043    & 8.278  & 4.766 \\
                       & TD & 3.238 & 5.174   & 5.039    & 8.296 & 4.768  \\
 \multirow{2}{*}{492} & RD & 3.239  & 5.169   & 5.048 & 8.264   & 4.766 \\
                       & TD & 3.238  & 5.166 & 5.047  & 8.259 & 4.772    \\
  \multirow{2}{*}{977} & RD & 3.239  & 5.168  & 5.046 & 8.273   & 4.766  \\
                       & TD & 3.238 & 5.165  & 5.049 & 8.256  & 4.774  \\
  \bottomrule
\end{tabular}
\end{center}
\footnotesize{Values in this table were measured at room temperature without load.}
\end{table}
The uncertainties of the Reitveld fit are 10$^{-5}$ $\AA$, 10$^{-4}$ $\AA$ and 10$^{-3}$ $\AA$ for the $\alpha$-Zr phase, $\delta$-hydride and $\chi$-phase, respectively. Although every sample was tested under the same geometry, order of $\sim$10$^{-3}$ $\AA$ uncertainty were generated while switching the samples. The measured lattice constants for $\alpha$-Zr and $\delta$-hydride are in good agreement with literature values \cite{santisteban2010hydride,douglass1971}. The
observed $\chi$-phase has measured lattice constants smaller than values reported by Erwin \emph{et al.} (a$_0=5.3$ $\AA$, c$_0=8.75$ $\AA$), but is in good agreement with Sande and Bement's result (a$_0=5.079$ $\AA$, c$_0=8.279$ $\AA$)  \cite{vander1974investigation}.
All phases have lattice constants comparable in two directions (TD and RD). The formation of hydride rim/blister has small impact on lattice constant of the $\alpha$-Zr phase as also observed by Daum \emph{et al.} \cite{daum2009identification}, where hydride precipitation did not induce measurable lattice strain on $\alpha$-Zr.

\subsection{Lattice strain evolution}
In this work the conventional Rietveld refinement is used to measure macrostrain or continuum equivalent response for the three phases ($\alpha$-Zr, $\delta$-hydride and $\chi$-phase). It should be noted here that a better method which is closer to actual material physics was developed by Daymond, which is weighting peaks contribution to obtain a theoretical continuum strain \cite{daymond2004determination}. Unfortunately, using peak weighting method requires the information of texture index of each planes, which could not be measured under the experimental geometry in this work. Also, an average over five peaks are required while using peak weighting method, which generally is not the case for a minor phase. Thus the best workaround for determining the continuum equivalent response in our case is to use the conventional Rietveld method.

However, careful assessment is necessary before applying the conventional Rietveld method.
Daymond pointed out that if the X-ray absorption (the absorption term for determining X-ray diffraction intensity) is strongly dependent on the wavelength, contribution from certain peaks will be reduced which results in invalid lattice parameters while using the Rietveld method \cite{daymond2004determination}. Also, it is commonly known that the conventional Reitveld method does not count for anisotropic response of the polycrystal materials.
Despite these inherent imperfections, valid Reitveld results were still obtained for both untextured and textured h.c.p materials \cite{daymond2004determination,daymond1999use}. Since in this work the wavelength was a constant during the whole experiment session, the absorption term should not influence the contribution of diffractions peaks toward the lattice parameters. Therefore it is plausibly to conclude that using the Reitveld method in this work should give results equivalent to the macrostrain response of each phases.

Thermal expansion coefficients from 25 \si{\degreeCelsius} to 200 \si{\degreeCelsius} were estimated from variation of the lattice parameters obtained by Reitveld method and were summarized in Table \ref{tab:thermal_coeff}. Lattice parameter variation of the $\chi$-phase from RT to 200 \si{\degreeCelsius} is within the uncertainty thus it is excluded from the table.  The estimated thermal expansion coefficients of $\alpha$-zirconium are in good agreement with the literature values (5.5$\times10^{-6}$ \si{\per\degreeCelsius} along the a-axis and 10.8$\times10^{-6}$ \si{\per\degreeCelsius} along the c-axis) \cite{skinner1953thermal}.
The estimated thermal expansion coefficient of $\delta$-hydride in this study is an order of magnitude smaller than reported by Yamanaka (2.7$\times10^{-5}$ \si{\per\degreeCelsius}) \cite{yamanaka1999thermal}, where pure (99.9\%) hydride was studied. A possible explanation is that the hydride expansion was constrained by the zirconium matrix. However, an earlier work by Kempter \emph{et al.} \cite{kempter1960thermal} reported a comparable value (2.5-3$\times10^{-6}$ \si{\per\degreeCelsius}) to this study.

\begin{table}[b]
\begin{center}
\caption{Summary of estimated thermal expansion coefficients.}
\label{tab:thermal_coeff}
\begin{tabular}{ccccccc}
  \toprule
  H Content & Direction & \multicolumn{2}{c}{$\alpha$-Zr} & $\delta$-hydride \\
  \cmidrule{3-4}

  (Wt.ppm)       &       & a-direction$\times10^{-6}$ (\si{\per\degreeCelsius}) & c-direction$\times10^{-6}$ (\si{\per\degreeCelsius})  &  a-direction$\times10^{-6}$ (\si{\per\degreeCelsius})   \\
  \midrule
  \multirow{2}{*}{98}  & RD &6.53  &5.95    &5.58  \\
                       & TD &6.2  &9.07    &6.22   \\
 \multirow{2}{*}{492} & RD & 6.26  & 7.3   & 5.34 \\
                       & TD & 6.28  &8.08   & 3.56    \\
  \multirow{2}{*}{977} & RD &5.7 &7.4  &4.08   \\
                       & TD &6.56 &8.24    &2.3   \\
  \bottomrule
\end{tabular}
\end{center}
\end{table}

The lattice strain of the three phases at 200 \si{\celsius} were determined from the relationship,
\begin{equation}
\label{eq:lattice strain}
\varepsilon_{\text{a}}=\frac{a_{\sigma}^{200 \si{\celsius}}-a_0^{200 \si{\celsius}}}{a_0^{200 \si{\celsius}}} \quad \text{or} \quad \varepsilon_{\text{c}}=\frac{c_{\sigma}^{200 \si{\celsius}}-c_0^{200 \si{\celsius}}}{c_0^{200 \si{\celsius}}}
\end{equation}

where the lattice parameters (a or c) are measured at 200 \si{\celsius} ($a_0^{200 \si{\celsius}}$ and $c_0^{200 \si{\celsius}}$) with small applied stress ($\sim3$ MPa for snugging samples) used as the references.
$a_{\sigma}^{200 \si{\celsius}}$ and $c_{\sigma}^{200 \si{\celsius}}$ are lattice parameters at 200 \si{\celsius} with a known applied stress $\sigma$.
It is worth to note that the c/a ratio was not constrained during the refinement, since the stiffness in the c and a directions are different and would experience different strain. Because $\alpha$-Zr and $\chi$-phase have a and c lattice parameters, lattice strains of these two phases were averaged according to multiplicities using \cite{daymond1999use}
\begin{equation}
\label{eq:ave_hcp_lattice strain}
\frac{2\varepsilon_{\text{a}}+\varepsilon_{\text{c}}}{3}
\end{equation}

An example of lattice strain evolution as a function of applied stress for the 977 wt.ppm sample is shown in Fig. 8. Uncertainties of strain are propagated from the Rietveld fitting, generating around $\pm$15, $\pm$80 and $\pm$300 micro-strain for $\alpha$, $\delta$ and $\chi$-phase, respectively.  Similar analysis was performed for the 492 and 98 wt.ppm hydrogen samples, with similar strain versus applied load trends and strain magnitudes, which is independent of the hydrogen content. Fig. 7 shows the strain measured on the three diffraction peaks of the $\delta$ hydride phase as well as the strain calculated from Rietveld method, suggesting that $\delta$-(111) will be the suitable plane for macro-strain profiling when it is impractical to use Reitveld or other weighting methods.
\begin{figure}[!ht]
\centering
\label{fig:delta_individualpeaks}
\includegraphics[scale=0.6]{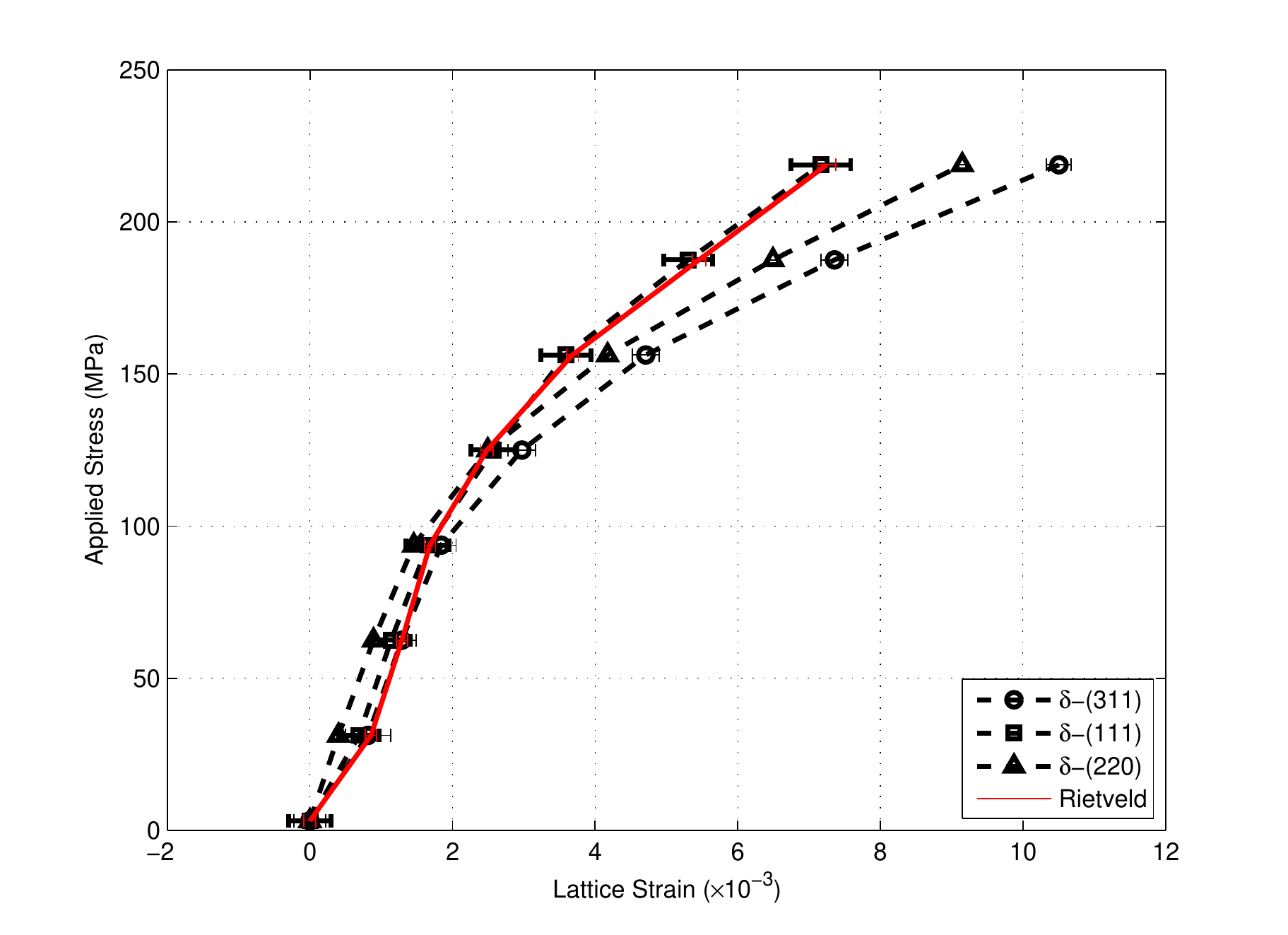}
\caption{$\delta$ hydride phase lattice strains response measured on the three individual diffraction planes and from Rietveld method at 200 \si{\celsius}}.
\end{figure}
\begin{figure}[!ht]
\centering
\label{fig:s25_strain_RD}
\includegraphics[scale=0.7]{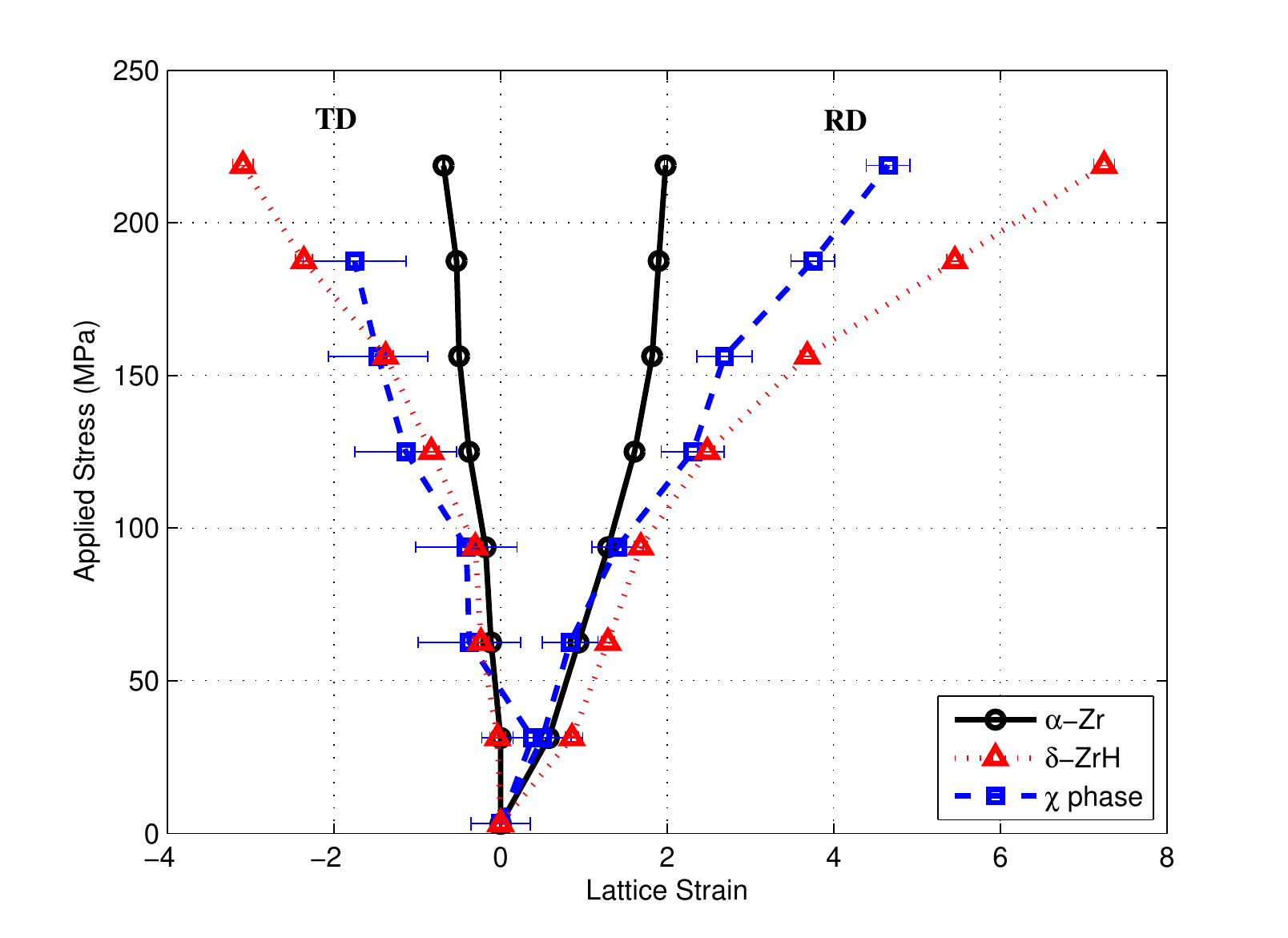}
\caption{Lattice strains evolution of three phases as a function of applied stress for the 977 wt.ppm sample. Sample was loaded axially along the RD. TD is perpendicular to the direction of the applied load. }
\end{figure}
\begin{table}[b]
\begin{center}
\caption{Summary of tension test results in the loading axial from the three phases.}
\label{tab:E}
\begin{tabular}{ccccccc}
  \toprule
  H Content &\multicolumn{3}{c}{Elastic Modulus (GPa)} & \multicolumn{3}{c}{Inflection Stress (MPa)} \\
  \cmidrule{2-5}
  \cmidrule{5-7}
  (Wt.ppm) & $\alpha$-Zr & $\delta$-hydride & $\chi$-phase               &$\alpha$-Zr & $\delta$-hydride & $\chi$-phase   \\
  \midrule
  \multirow{1}{*}{98}& 91.9$\pm$2.8 & 66.8$\pm$24.4  & 85.9$\pm$41.2    & 99.8$\pm$23.7  & 115.9$\pm$20.2 & 133.5$\pm$71.3\\

  \multirow{1}{*}{492} & 80.5$\pm$7.5  & 66$\pm$6.9   & 81.9$\pm$34.7 & 101.8$\pm$23.3   & 105.3$\pm$9.4& 118.8$\pm$70.1 \\

  \multirow{1}{*}{977} & 90.6$\pm$2  & 75.6$\pm$10.3  & 66.6$\pm$21.8 & 138.8$\pm$38.2   & 119.3$\pm$12.9 & 103.8$\pm$76.7  \\

  \bottomrule
\end{tabular}
\end{center}
\footnotesize{$^*$ NA is for data which is masked by large measured error.}
\end{table}

For all three phases, two regions are qualitatively distinguished from the lattice strain behavior shown in Fig. 8.  The first is the initial linear elastic region where the strains of three phases increase linearly with increased stress, which is qualitatively defined as $\sigma_{\text{applied}}<90$ MPa for the $\delta$ and $\chi$-phase, and as $\sigma_{\text{applied}}<125$ MPa for the $\alpha$-Zr phase. Elastic modulus values along the loading axial (RD) are measured from the data in region I using linear least square fitting with York method (propagating uncertainties from each data points) \cite{york1966least}, and are reported in Table \ref{tab:E}. It is noticeable that the uncertainty of strain data points ($\pm$300 micro-strain) from the $\chi$-phase propagate around $\pm30\%$ uncertainty in measured elastic modulus value. This is due to weak diffraction reflection from the minor $\chi$-phase even when the refinement Fig. \ref{fig:fit_chi} shows a reliable result. Therefore, this modulus value should be treated with caution.
The measured elastic modulus of $\delta$-hydride phase is similar to the value of $\alpha$-Zr phase, both are in good agreement with literature values at 200 \si{\celsius} \cite{puls2005experimental,douglass1971}.

The second region is denoted by load partitioning, with significant lattice strains observed in the $\delta$-hydride and $\chi$ phases and an invariant response in the $\alpha$-Zr matrix. The transition between region I and II is caused by load transfer from the zirconium matrix to the other two phases,  which is confirmed using finite element method by Kerr \emph{et al.} \cite{kerr2008strain}. Load transfer to the plastically stronger second phase is generally believed to be responsible for the post-transition matrix yield \cite{kerr2008strain,young2007load}. In this case, the elastic strain in the Zry-4 matrix becomes invariant with increasing applied load beyond the transition. The transition therefore signals the transfer of applied load to the stronger $\delta$-hydride and $\chi$ phase.
Since the load partitioning region is a subsequent linear region of region I (linear elastic region) with a decreased slope, the inflection point of the two regions can be estimated using two linear fits and determining the intersection between the two lines, as shown in Fig. (9). The inflection points are represented as magnitude of the applied stress and are reported in Table \ref{tab:E}.
\begin{figure}[!ht]
\begin{center}
\label{fig:transfit}
\subfigure[] {
\includegraphics[scale=0.27]{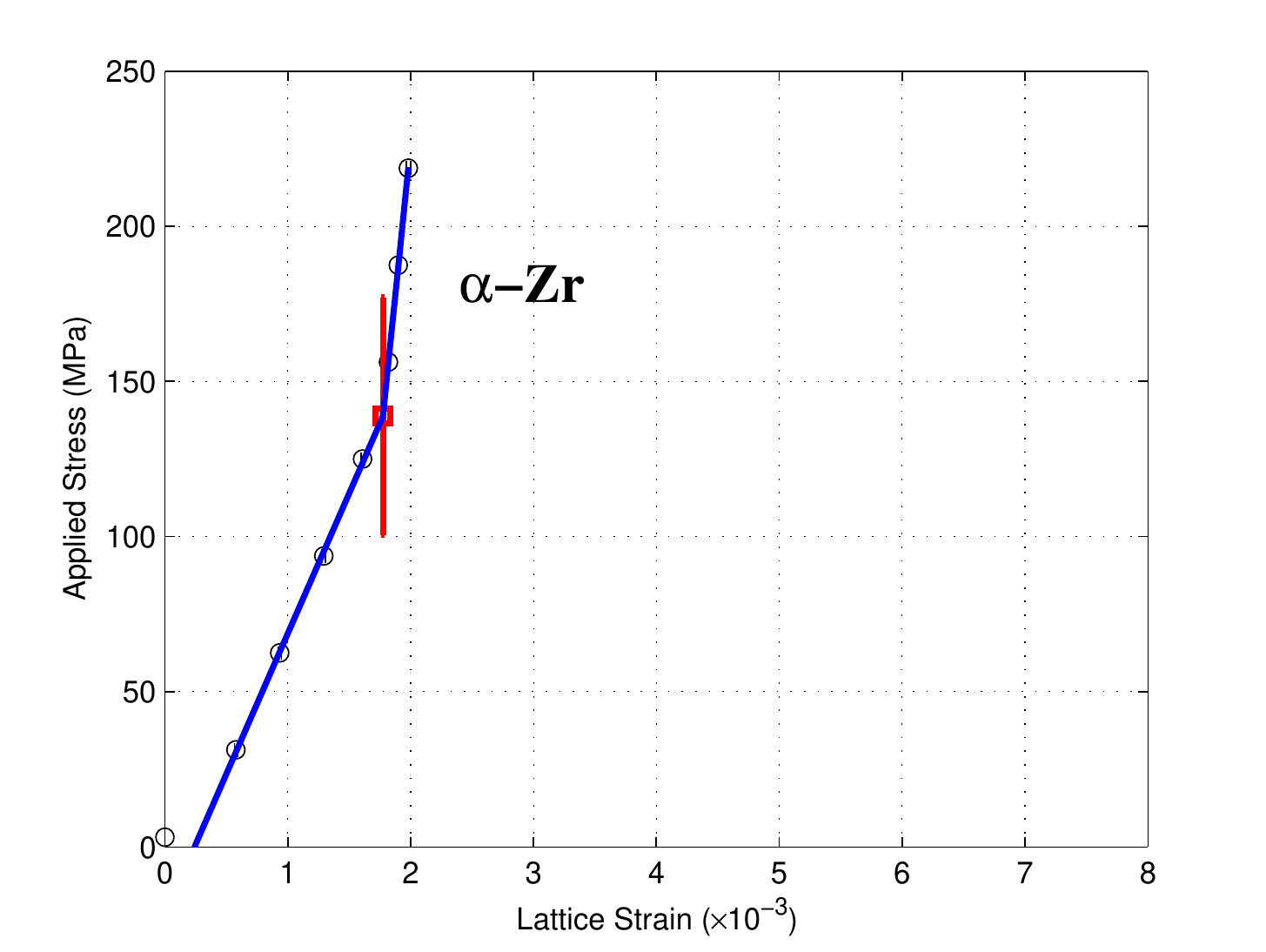}}
\subfigure[] {
\includegraphics[scale=0.27]{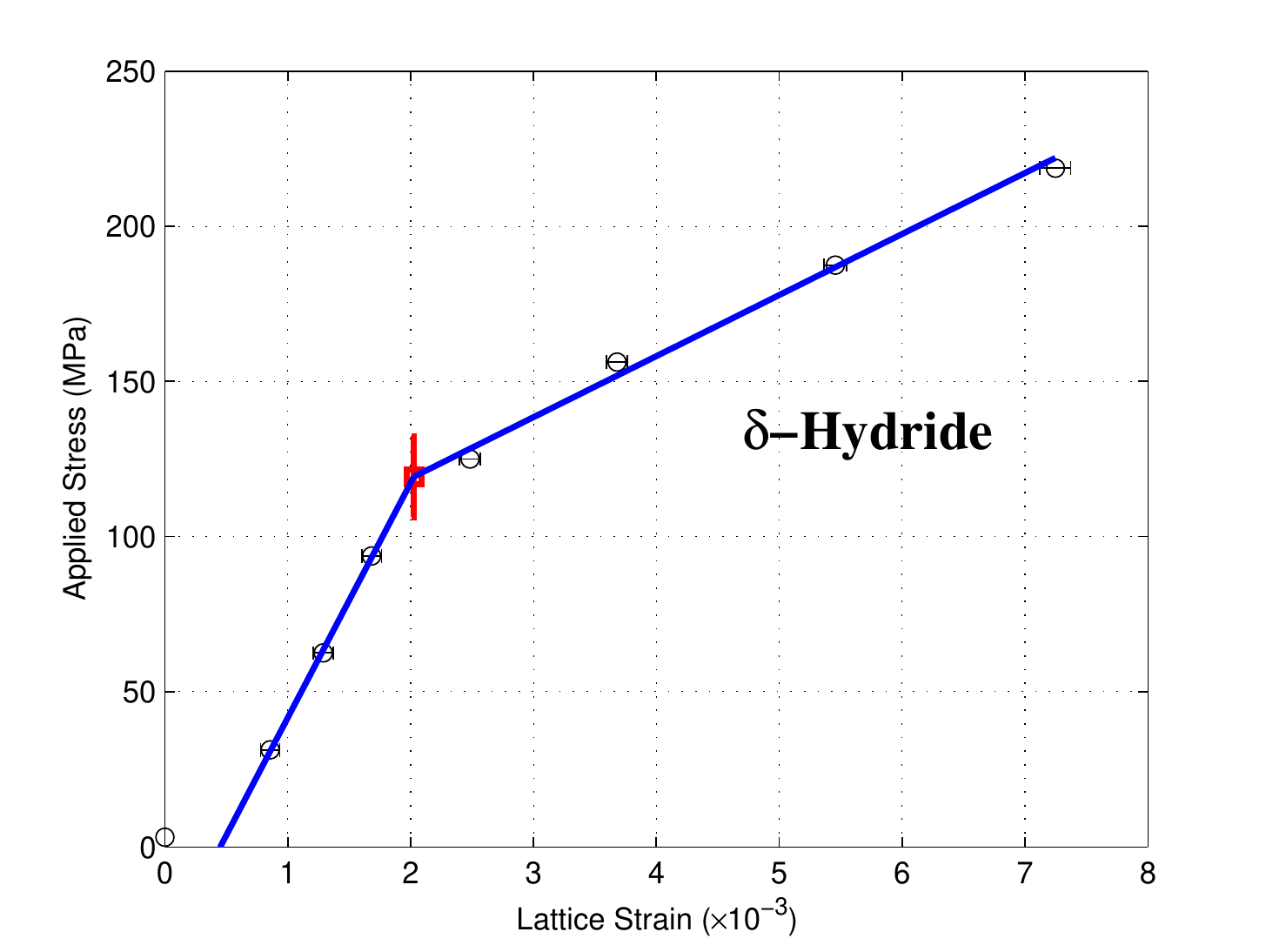}}
\subfigure[] {
\includegraphics[scale=0.27]{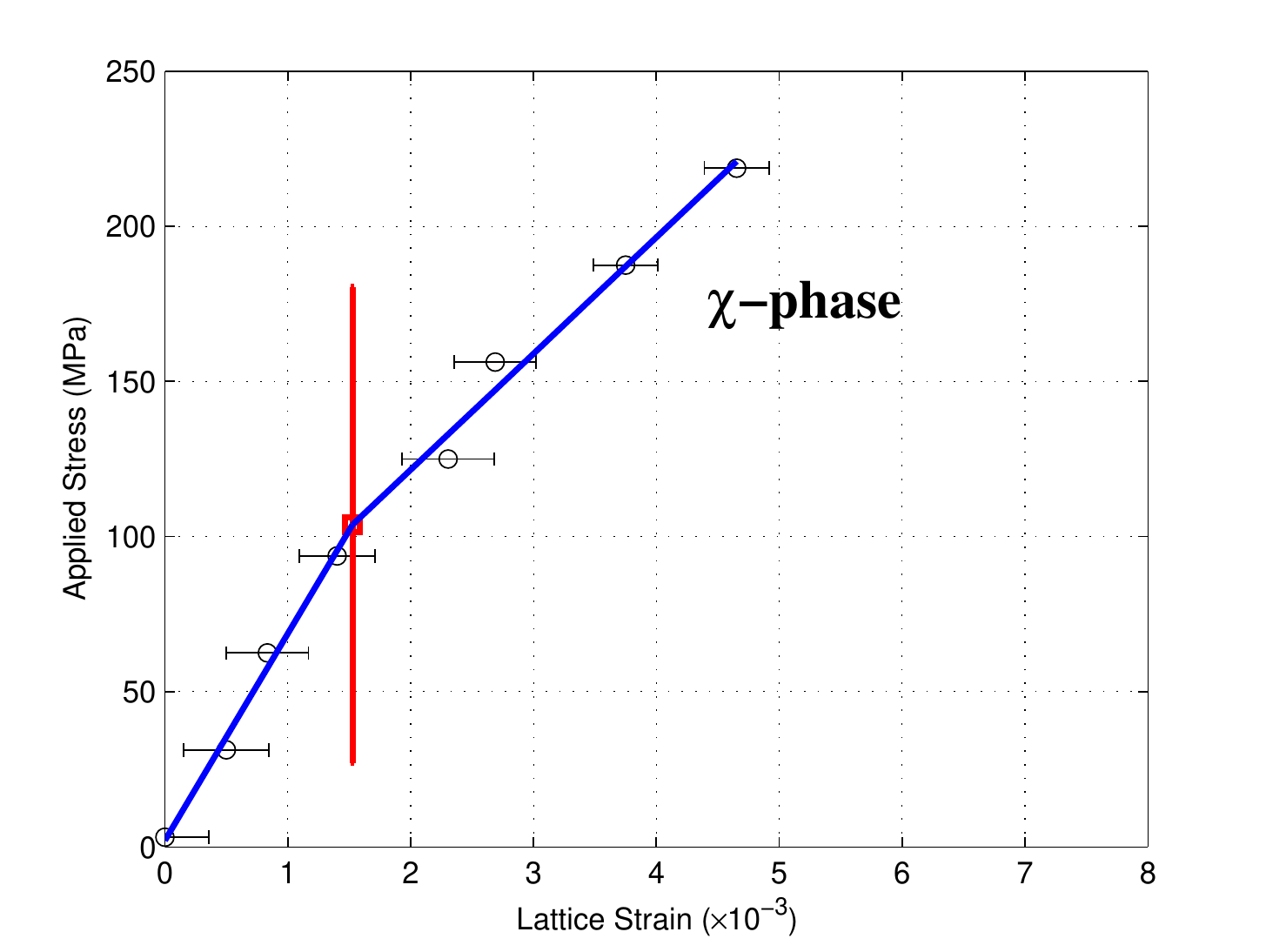}}
\caption{An example of inflection stress determination from the data along the loading axial (RD) for the 977 wt.ppm sample. (a) $\alpha$-Zr. (b) $\delta$-hydride. (c) $\chi$-phase. Data were taken at 200 \si{\degreeCelsius}.}
\end{center}
\end{figure}
The large uncertainty of the $\chi$-phase is again propagated from the large lattice strain uncertainty. If the determination of inflection stresses did not inherit the uncertainties of the lattice strains, the errors for the inflection stresses will be within 10\% and the three phases will have  comparable inflection stresses.

There is no measurable lattice strain for all three phases after the short-term creep tests. Although creep may occurred plastically (by diffusion), it was not revealed by diffraction in this case.
\subsection{Load partitioning between the Zr and the hydride phase}
Since the measured lattice strains from the three phases should represent the average response of each phase in the material, the three principal stresses can then be calculated using
\begin{subequations} \label{vonmises2}
\begin{equation}
\sigma_{11}=\frac{E}{1+\nu}\varepsilon_{11}+\frac{\nu E}{(1+\nu)(1-2\nu)}(\varepsilon_{11}+\varepsilon_{22}+\varepsilon_{33})
\end{equation}
\begin{equation}
\sigma_{22}=\sigma_{33}=\frac{E}{1+\nu}\varepsilon_{22}+\frac{\nu E}{(1+\nu)(1-2\nu)}(\varepsilon_{11}+\varepsilon_{22}+\varepsilon_{33})
\end{equation}
\end{subequations}
if assumed the elastic anisotropy between the ND and TD is small. In these equations $11$, $22$ and $33$ corresponds to the RD (loading direction), TD and ND, respectively. Stresses along the TD and ND are the Poisson response to the applied tensile load.
$\varepsilon_{11}$, $\varepsilon_{22}$ and $\varepsilon_{33}$ are the three measured principal lattice strain. $E$ is the elastic modulus listed in Table \ref{tab:E} and $\nu$ is the Poisson's ratio of each phase determined from lattice strain data in region I using $\nu=\lvert E_{11}/E_{22} \rvert$.

\begin{figure}[!ht]
\centering
\includegraphics[scale=0.6]{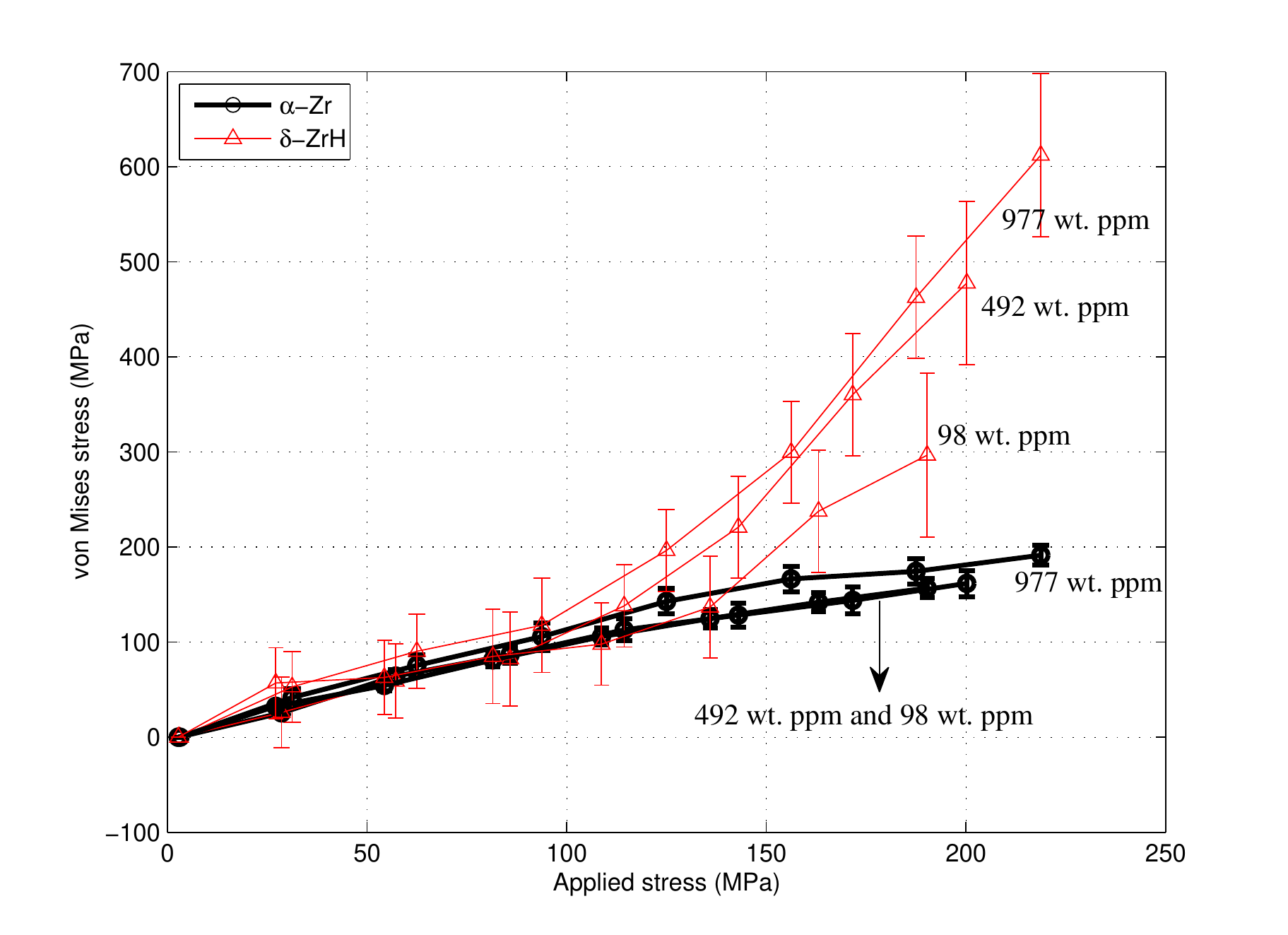}
\caption{The evolution of von Mises stress as a function of the applied stress at 200 \si{\degreeCelsius}. The $\alpha$-Zr phase is represented as the black solid line and the red solid line represents the $\delta$-hydride phase.}
\label{fig:vonmises}
\end{figure}

The amount of load transferred in the load partitioning region can then be quantified by calculating the average von Mises stress ($\sigma_{eff}$) of each phase \cite{shen20023d}, using
\begin{equation} \label{vonmises}
\sigma_{eff}=\sqrt{\frac{(\sigma_{11}-\sigma_{22})^2+(\sigma_{22}-\sigma_{33})^2+(\sigma_{33}-\sigma_{11})^2}{2}}
\end{equation}
. This analysis does not include the $\chi$-phase due to the large uncertainty of this relative minor phase. An example of the evolution of the von Mises stress as a function of the applied stress is shown in Fig. \ref{fig:vonmises}. The magnitude of von Mises stresses on the two phases only slightly varied between the three samples.
The evolution of the von Mises is very similar to the lattice strain response and two distinct regions can again be identified.  The von Mises stresses in both phases increased linearly with applied stress in a nearly identical manner before the transition.  The observed transition is again associated with load partitioning and a lower (and invariant) effective stress within the $\alpha$-Zr matrix.  A significantly higher von Mises stress within the hydride phase is observed post-transition.  The effective or von Mises stress in the hydride phase exceeds $400$ MPa at an applied stress of $200$ MPa.  Load transfer behavior is desired in composite materials since it represents a shift of load to stronger particles/fibers embedded in a softer matrix  \cite{young2007load,mo2014synchrotron}.  However, in this case, load partitioning will likely facilitate brittle failure of the hydride phase.  It is therefore important to quantify the transition stress and effective stress magnitude within the $\delta$-hydride phase. Since the internal stress significantly affects the behavior of zirconium hydride phase (Ex: increases the stress also increases the terminal solid solubility for hydride precipitation), evaluating the von Mises stress on each phase help better understanding and predicting the cladding performance during the fuel rod life-cycle.

\section{Conclusion}
High energy synchrotron x-ray diffraction was utilized to study the strain response of $\alpha$-Zr, $\delta$-hydride and \emph{C-14} $\chi$ phases in a CWSR Zry-4 plate with hydride blister/rim distribution with an external tensile load applied along the RD at 200 \si{\celsius}. Findings are summarized below:
\begin{enumerate}
\item The hydride in the blister/rim structure has similar texture relationship with the $\alpha$-Zr as the uniformly distributed hydrides.
\item For overall hydrogen content $\leq$ 977 wt.ppm, the f.c.c $\delta$ hydride is the only hydride phase identified in the specimens by the synchrotron X-ray diffraction.
\item The conventional Rietveld refinement suggests that $\delta(111)$ is a suitable plane for macrostrain profiling for the $\delta$-hydride phase.
\item From strain-stress curves of the three phases, a linear and load partitioning regions can be identified.  The lattice strain of hydride phase increases significantly while the lattice strain of $\alpha$-Zr becomes invariant in the load partitioning region.  The measured elastic modulus of the three phases from the linear elastic region are similar at 200 \si{\celsius}.
\item The von Mises stresses in the three phases can be quantified using the average strain obtained by the Rietveld method.






\end{enumerate}
\section{Acknowledgments}
The authors would like to acknowledge Y. Miao and M.S. Elbakhshwan for their assistance with the experimental work. The author also greatly acknowledge M. Kun for fruitful discussion on data analysis. The assistance of M. Sardela , J. Mabon, and T. Shang on X-ray, SEM, and Sputter Coater instrument at the Frederick Seitz Materials Research Laboratory Central Research Facilities, Univ. of Illinois, are also gratefully acknowledged. This work was supported by the US Department of Energy Nuclear Energy Fuel Aging in Storage and Transportation under Grant No. IRP-2011-05352. The work was also carried out in part in the Frederick Seitz Materials Research Laboratory Central Research Facilities, University of Illinois, which are partially supported by the US Department of Energy under Grants DE-FG02-07ER46453 and DE-FG02-07ER46471. Usage of the Advanced Photon Source was supported by the US Department of Energy, under Contract No. DE-AC02-06CH11357.
\clearpage
\bibliographystyle{model1a-num-names}
\bibliography{hydride_synchrotron}

\end{document}